Algorithmically generated subject categories based on citation relations:

An empirical micro study using papers on overall water splitting

Robin Haunschild*$, Hermann Schier*‡, Werner Marx*, & Lutz Bornmann**

$ Corresponding author

* Max Planck Institute for Solid State Research

Heisenbergstr. 1,

70569 Stuttgart, Germany.

E-mail: R.Haunschild@fkf.mpg.de

** Division for Science and Innovation Studies

Administrative Headquarters of the Max Planck Society

Hofgartenstr. 8,

80539 Munich, Germany.

E-mail: bornmann@gv.mpg.de

‡ Hermann Schier passed away before the final version of the manuscript has been written. He gathered and analyzed many of the results and contributed to an earlier draft of this manuscript.

**Abstract**

One important reason for the use of field categorization in bibliometrics is the necessity to make citation impact of papers published in different scientific fields comparable with each other. Raw citations are normalized by using field-normalization schemes to achieve comparable citation scores. There are different approaches to field categorization available. They can be broadly classified as intellectual and algorithmic approaches. A paper-based algorithmically constructed classification system (ACCS) was proposed which is based on citation relations. Using a few ACCS field-specific clusters, we investigate the discriminatory power of the ACCS. The micro study focusses on the topic "overall water splitting" and related topics. The first part of the study investigates intellectually whether the ACCS is able to identify papers on overall water splitting reliably and validly. Next, we compare the ACCS with (1) a paper-based intellectual (INSPEC) classification and (2) a journal-based intellectual classification (Web of Science, WoS, subject categories). In the last part of our case study, we compare the average number of citations in selected ACCS clusters (on overall water splitting and related topics) with the average citation count of publications in WoS subject categories related to these clusters. The results of this micro study question the discriminatory power of the ACCS. We recommend larger follow-up studies on broad datasets.

# 1   Introduction

In bibliometrics, it is often necessary to compare the impact of publications from different fields[1]. However, it should be avoided to use bare citation counts ("times cited") from Web of Science (WoS, Clarivate Analytics) or Scopus (Elsevier) for such comparisons. Many bibliometric studies have shown that there are large differences in citation rates between fields, which cannot be explained by the quality of publications (see, e.g., Bornmann & Marx, 2015). Field-normalized indicators have been developed in bibliometrics which make cross-field comparisons possible. "The idea of these indicators is to correct as much as possible for the effect of variables that one does not want to influence the outcomes of a citation analysis, such as the field … of a publication" (Waltman, 2016, p. 375). The use of normalized indicators in research evaluation is one of the guiding principles for research evaluation in the Leiden manifesto for research metrics (Hicks, Wouters, Waltman, de Rijcke, & Rafols, 2015).

In recent years, several methods have been proposed for the calculation of normalized citation scores. An overview of these methods can be found, for example, in Mingers and Leydesdorff (2015), Waltman (2016), and Bornmann and Marx (2015). Today, indicators based on the idea of counting highly cited publications are seen as a robust method for measuring citation impact across fields (Wilsdon et al., 2015). An important topic in the calculation of field-normalized indicators is the way in which research fields are defined, i.e. which field-categorization schema is used in bibliometrics to calculate the expected number of citations (Wilsdon, et al., 2015).

The most common approach in bibliometrics is to work with subject categories defined by Clarivate Analytics in WoS or by Elsevier in Scopus. These subject categories are based on sets of journals publishing research from similar areas. However, the use of journal sets for field-normalization is heavily criticized. The most critical point is papers published in multi-disciplinary journals which cannot be assigned to corresponding fields using journal sets (Hui, 2015; Kronman, Gunnarsson, & Karlsson, 2010). Alternative approaches which can be used instead of journal sets have been classified by Wang and Waltman (2016) in mono-disciplinary and multi-disciplinary classification systems.

---

[1] We use the terms "field" and "topic" interchangeably, because the distinction between the two is rather arbitrary (Sugimoto & Weingart, 2015).

A mono-disciplinary classification system "covers publications in one particular research area and usually provides a classification at a relatively high level of detail" (Wang & Waltman, 2016, p. 348). Mono-disciplinary classification systems, as the Physics and Astronomy Classification Scheme (PACS, see https://publishing.aip.org/publishing/pacs/pacs-2010-regular-edition) system used in this study, are mostly expert-based approaches (Wang & Waltman, 2016) where experts in the fields (at least the authors of a paper) assign papers to corresponding subject categories. Nowadays, paper classification of PACS is supported by machine-indexing but expert controlled. PACS is included in the broader classification scheme of the INSPEC database. At the highest hierarchical level, INSPEC features the sections A (Physics), B (Electrical Engineering & Electronics), C, (Computers & Control), and D (Information Technology) (The Institution of Electrical Engineers, 1992). The section A of INSPEC is identical with PACS.

Waltman and van Eck (2012) introduced a method for algorithmically constructing classification systems (ACCS) at the level of individual publications. The method is a multi-disciplinary classification system and is based on citation relations between publications. The approach which is explained in more detail in section 2 plays a prominent role among the available schemes, because it is used in the Leiden ranking (a university ranking based on bibliometrics, available at http://www.leidenranking.com/) for the calculation of field-normalized impact scores (the Leiden ranking uses a different solution of the ACCS than the system studied, here). The method employed by Waltman and van Eck (2012) uses direct citation relations between papers for classification. They provide software referred to as "Modularity Optimizer" which uses direct citations as similarity measure. The general ACCS concept can be used, however, with different similarity measures (e.g., bibliographic coupling, co-citations, and textual comparison).[2] The results of Klavans and Boyack (2017) show in general that classification systems based on journal schemes are inaccurate compared to algorithmically constructed classifications. Sjögårde and Ahlgren (2018) discuss "how the resolution parameter given to the Modularity Optimizer software can be calibrated so that obtained publication classes correspond to the size of topics" (p. 149).

In this case study, we investigate the ability of the method to reliably assign publications to fields. This study is not intended to undertake a broad comparison between ACCS and other field classification systems, but to analyze one specific field, namely, "overall water splitting", in more detail. The use of this field has three advantages: (1) Most of the publications can be reliably compiled in WoS by a topic search. (2) One of the labels for a cluster in ACCS is "overall water splitting" (cluster 3.7.3). (3) Three of the four authors of this paper have a background in chemistry and physics. Therefore, we are able to provide a qualitative perspective on the search results. Research on overall water splitting is important for hydrogen gas production from water. The direct water splitting using solar cells or other renewable energy sources is especially appealing. Such a detailed and qualitative approach is not possible on a large scale, like the study by Klavans and Boyack (2017).

In the empirical part of this study, we present the results of several analysis steps: (1) Experts examined (read) a sample of papers in ACCS cluster 3.7.3 to determine whether they really deal with the topic "overall water splitting"? (2) We investigate the spread of publications found by the WoS topic search over the ACCS clusters: Are most of the "overall water splitting" publications assigned to cluster 3.7.3? (3) We take the other way around and study the spread of publications in the ACCS cluster "overall water splitting" over WoS and PACS subject categories (SCs). (4) We compare the ACCS cluster 3.7.3 with related clusters of similar size (3.7.2 and 3.7.4) to investigate the discriminatory power of the ACCS. Papers assigned to different clusters should differ in terms of content. (5) We study citation impact differences of the papers in these related clusters of similar size. The clusters on the same hierarchical level are ordered by the number of papers in the cluster.

---

[2] For the sake of brevity, we refer to the classification system studied here as ACCS.

## 2      Field classification systems used in this study

Science is structured by disciplines (e.g. physics or chemistry), whereby each discipline is a specific domain of particular research traditions including paradigms, codes of practice, and methods (Ziman, 1996). Although it is practically impossible that a scientist is not located in at least one discipline, the disciplines are rather loosely organized – as an "invisible college" (Andersen, 2016). Scientific publications are the main research outcome of scientists. The loosely organized disciplines might be one of the main reasons why there does not exist an established and widely accepted field categorization scheme for scientific papers. The two most important multidisciplinary literature databases used for bibliometric purposes (e.g. field normalization) are the WoS and Scopus (Wang & Waltman, 2016; Wouters et al., 2015). The around 250 WoS SCs (such as biochemistry or condensed matter physics) are based on journal sets. Thus, each paper in WoS is assigned to one category or more based on the assignment of the publishing journals to the WoS SC. The problems of using the WoS SC for field-normalization in bibliometrics are explained by Haddow and Noyons (2013) in detail.

PACS is a mono-disciplinary classification system which was developed by the American Institute of Physics (AIP). PACS classifications are assigned to papers by the authors themselves. According to Radicchi and Castellano (2011) "this guarantees an optimal classification into fields, overcoming the nontrivial problem of attributing, a posteriori, papers to fields. PACS codes are composed of three fields *XX.YY.ZZ*, where the first two are numerical (two digits each) and the third is alphanumerical. For our purpose we consider only

the first digit of the *XX* code, which provides a classification into very broad categories". Other classification systems of professional databases are Chemical Abstract (CA) sections offered by the Chemical Abstract Service (CAS) (Bornmann & Daniel, 2008; Bornmann, Marx, & Barth, 2013) and Medical Subject Heading (MeSH) terms by the United States National Library of Medicine (Leydesdorff & Opthof, 2013).

The ACCS developed by Waltman and van Eck (2012) is based on a transparent clustering technique which assigns papers to field-specific clusters based on direct citations between single papers. The algorithm needs three basic parameters as input besides the direct citation network: (1) the number of levels of the classification system. Using only one level results in a non-hierarchical classification system. (2) The resolution parameter determines how much detail is offered at each level. The resolution parameter is bound between 0 and 1. (3) The minimum number of publications per cluster needs to be specified, too. The latter two parameters can have different values for each level of the classification system.

The ACCS has four important advantages: (1) The classification works on the level of single publications and not journals (like the WoS scheme). Thus, the assignment of publications to fields is more detailed and difficulties with multidisciplinary journals are avoided. (2) Each paper in the literature database is assigned to a field only once. Usually, field classification systems (e.g. WoS and Scopus journal sets or expert-based systems, such as PACS and CA sections) assign papers to more than one field which complicates the statistical analysis and the calculation of field-normalized indicators because different counting methods can be applied, e.g., fractional counting, (scaled or unscaled) full counting, and multiplicative counting (Haunschild & Bornmann, 2016). (3) The ACCS is not restricted to a single discipline (such as PACS). Thus, it can be applied to the entire literature database (WoS). (4) The ACCS methodology for the clustering of publications in literature databases is freely available. Thus, the ACCS can be implemented in every in-house literature database which includes direct citation links between publications.

The ACCS methodology is based solely on direct citation links between publications. Thus, frequently used other techniques, such as co-citations, bibliographic coupling, shared words in titles or abstracts, are not part of this methodology. Since there are large differences in citation density between fields (Marx & Bornmann, 2015), Waltman and van Eck (2012) corrected these differences by normalizing relatedness scores by fractional citation counting. Within the ACCS approach of clustering papers, a large-scale optimization problem was solved by introducing the so-called smart local moving algorithm (Waltman & Eck, 2013). This approach is able to handle very large datasets: In the first application of their approach, Waltman and van Eck (2012) classified many millions of papers from the sciences and social sciences published between 2001 and 2010. The received classification system distinguishes between three granularity levels with a minimum of 120,000, 5,000, and 50 papers per cluster. The three levels are hierarchically ordered in the sense that level 1 clusters are nested in level 2 clusters which are themselves clustered in level 3. In such a classification system with three levels, each cluster will be referred to as X.Y.Z where X, Y, and Z are natural numbers. In non-hierarchical classification systems, only one natural number is used to refer to an individual cluster.

Although the ACCS offers many advantages compared to other existing classification systems, it has been criticized. Leydesdorff and Milojević (2015) summarize the critique as follows: "Because these 'fields' are algorithmic artifacts, they cannot easily be named (as against numbered), and therefore cannot be validated. Furthermore, a paper has to be cited or contain references in order to be classified, since the approach is based on direct citation relations. However, algorithmically generated classifications of journals have characteristics very different from content-based (that is, semantically meaningful) classifications … The new Leiden system is not only difficult to validate, it also cannot be accessed or replicated from outside its context of production in Leiden" (p. 201).

# 3    Datasets used

In this study, we used three datasets:

(1) Waltman and van Eck (2012) have provided labels for the ACCS clusters. One of the labels for cluster 3.7.3 is "overall water splitting". The other labels are "bivo4", "solid state reaction method", "sacrificial agent", and "photocatalytic h". $BiVO_4$ seems to occur often enough in titles and abstracts of papers in this cluster to appear as one of the labels. However, chemical compounds are not useful field classifications because they are studied in different fields with different foci. The label "photocatalytic h" is rather redundant to "overall water splitting" as water splitting is usually performed via photocatalytic techniques. The labels "solid state reaction method" and "sacrificial agent" are also not helpful because these terms are too broad to use them for field classification. We were able to match all papers (n=1739) of the cluster 3.7.3 via the WoS UTs with their WoS SCs, but only 686 (39.4%) papers could be matched via the DOI with their INSPEC categories in STN (an online database for physics and related areas, see http://www.stn-international.de). The INSPEC classification system was reduced to the second level to be comparable with the ACCS. Papers with more than one classification are counted multiple times in the WoS and INSPEC schemes.

(2) We used the WoS search query 'TS=(( (overall OR photocataly* OR cataly* OR "visible light") SAME ("water splitting" OR "splitting of water")) OR (("hydrogen evolution" OR "h-2 evolution" OR "hydrogen production" OR "h-2 production" OR "h-2 generation" OR "hydrogen generation") NEAR/4 (water OR "visible light" OR photocataly*))) AND PY=2001-2010' within the indices Science Citation Index Expanded and Conference Proceedings Citation Index- Science. Essentially, the query has two parts: the first part captures publications dealing with photocatalytic water splitting, and the second part captures publications dealing with hydrogen evolution from water using a photocatalyst. Both publication sets deal with the topic of "overall water splitting" but the wording is different in title, abstract, and keywords. We refined the resulting publication set to the document types article, letter, and review and excluded the following WoS SCs: "Biochemistry & Molecular Biology", "Biophysics", "Plant Sciences", "Biotechnology & Applied Microbiology", "Biology", "Water Resources", "Cell Biology", and " Microbiology ". This topic search yields most of the articles, reviews, and letters from the research on direct water splitting by solar energy. The excluded SCs (69 papers) mainly include biological water splitting and wastewater treatment or other false positives. At the date of search (January 24$^{th}$, 2018), we found 1706 records in the WoS. For 1692 of the records, we were able to match them with the ACCS clusters via the WoS UTs. We downloaded the classification system which was described in the previous section from http://www.ludowaltman.nl/classification_system/ on November 7$^{th}$, 2014.

(3) We compare the papers of cluster 3.7.3 with the related clusters of similar size 3.7.2 (n=2645 papers) and 3.7.4 (n=1677 papers). The labels of cluster 3.7.2 are "n doped tio2", "n doping", "nitrogen", "tio2 lattice", and "tio2 xnx" and the labels of cluster 3.7.4 are "catalyst loading", "initial dye concentration", "operational parameter", "azo dye", and "decolorization". The labels of cluster 3.7.2 seem rather redundant. We would expect papers about $TiO_2$, its lattice, and nitrogen doping of $TiO_2$ in this cluster which is only partly a useful field classification because such aspects are investigated in different scientific fields (e.g., physical chemistry, analytic chemistry, condensed matter physics). The labels of cluster 3.7.4 indicate that this cluster contains papers about azo dyes and catalysis. All papers from the related clusters could be matched via the WoS UTs with their WoS SC. Also, for all papers in the three clusters the citations are determined (from publication year until the end of 2014).

# 4 Results

## 4.1 Intellectual investigation of papers in ACCS cluster 3.7.3

We checked by two chemists (RH and WM) to which extent the ACCS cluster 3.7.3 contains papers regarding the topic "overall water splitting". Since the chemists are not able to inspect large paper sets, we drew a random sample (n=123) from the n=1739 papers in ACCS 3.7.3 cluster "overall water splitting". We undertook a power analysis to estimate the sample size, which is oriented towards the level of agreement between the two chemists (Reichenheim, 2001). The two chemists independently assigned a value of 0 (paper does not belong to the topic of "overall water splitting") and 1 (paper does belong to the topic of "overall water splitting") to the sample papers. The results are shown in Table 1.

Table 1: Results of the manual inspection of a random sample of 123 publications from data set 2

|  | Chemist 1 | Chemist 2 |
|---|---|---|
| Number of papers which belong to the topic "overall water splitting" | 62 | 70 |
| Number of papers which do not belong to the topic "overall water splitting" | 61 | 53 |

Chemist 1 assigned 62 (50.4%) of the 123 papers to the topic of "overall water splitting". Chemist 2 assigned 70 (56.9%) of the 123 papers to the topic of "overall water splitting". The assignments agreed in 105 of the cases (85.4%). We additionally calculated a kappa coefficient which indicates with a value of 0.71 a "substantial agreement" according to Landis and Koch (1977). 57 (46.3%) of the sampled publications were identified to belong to the topic of "overall water splitting" by both chemists. Assuming that our random sample is representative for the ACCS cluster 3.7.3, one can estimate that nearly half of the publications in this cluster deal with the topic of "overall water splitting".

The random sample can also be used to validate the topic search (data set 2). Out of the 57 sampled publications with agreeing judgement by the two chemists (see above), 53 publications (93.0%) are included in the topic search using our WoS search query. 48 of the sampled publications were identified by both chemists to belong to a different topic. None of these 48 publications is included in the topic search. This indicates a rather high precision and recall of the topic search.

## 4.2 Comparison of WoS topic search and ACCS cluster

Figure 1 shows the distribution of the publications found by the WoS topic search (data set 2) across ACCS clusters. The 1692 publications are assigned within the ACCS to 248 different clusters, only the top 20 clusters comprising 1234 publications are shown in Figure 1. The blue bars show the absolute and the red dots the cumulative relative numbers of papers for each ACCS cluster. 41.1% (n=695) of the papers of our topic search were found in the cluster 3.7.3. The remaining 58.9% (n=997) of the publications are located in other clusters. The cluster 3.7.2 is the only cluster with at least 5% of the publications from the topic search. The cluster 3.7.2 contains 90 papers (5.3%) of the topic search. The rest of the 907 papers is spread across 246 other ACCS clusters with less than 5% of the papers of the topic search each.

The labels of the cluster 3.7.2 were already listed and discussed in the description of data set 3 in the methods section. The fact that the second highest number of publications from our topic search was found in cluster 3.7.2 indicates that some of the publications in clusters 3.7.3 and 3.7.2 may belong to a similar topic.

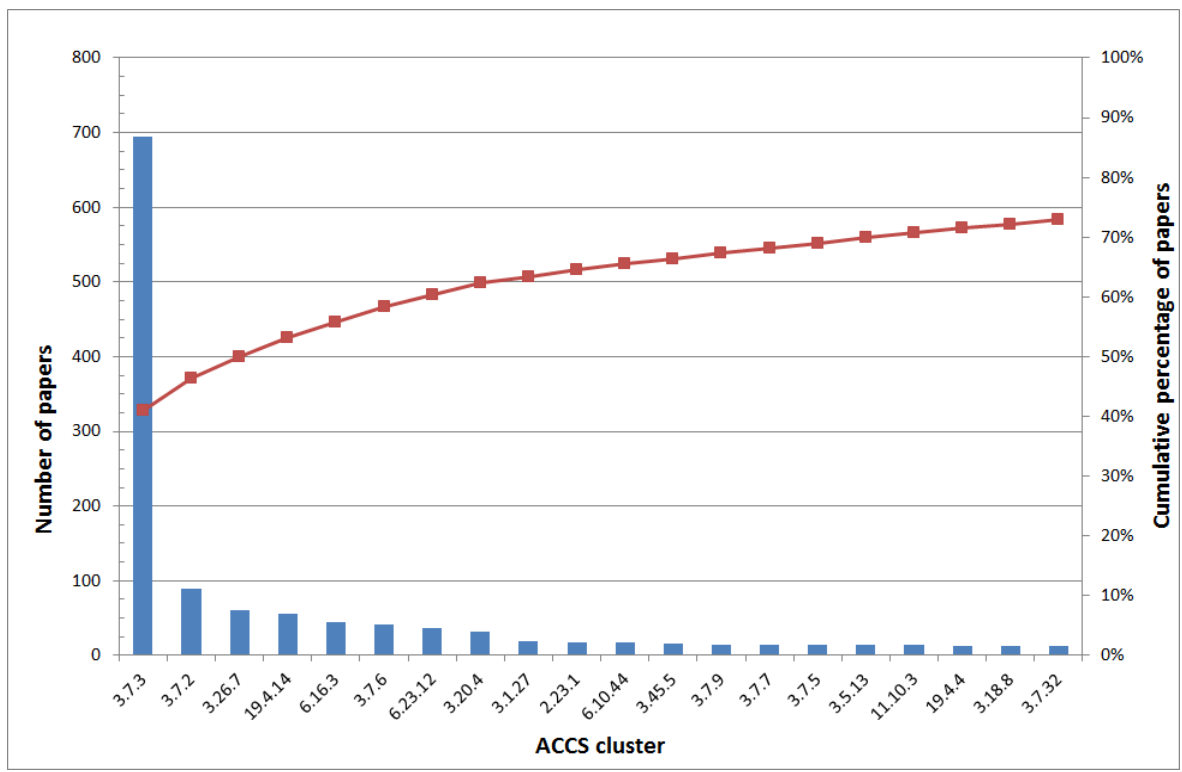

Figure 1: Distribution of the publications found by the WoS topic search (data set 2) across ACCS clusters. The blue bars show the absolute numbers of publications in ACCS clusters (y-axis on the left) and the red dots the cumulative relative numbers (y-axis on the right).

All papers from the topic search could be matched with WoS classifications, but only a subset of the papers with a DOI (n=806) could be matched with their INSPEC classifications via newSTN (www.stn.org). As each paper can belong to multiple WoS and INSPEC classifications, the sum exceeds the unique number of papers. The distribution of the papers found by the WoS topic search (data set 2) across INSPEC and WoS classifications is shown in Figure 2. One of the most important applications of the research field of overall water splitting, energy conversion and energy storage, appears as the second most populated SC in both classification systems ("Energy research and environmental science" in INSPEC and "Energy & Fuels" in WoS). The category labels of WoS and INSPEC classifications agree quite well with each other.

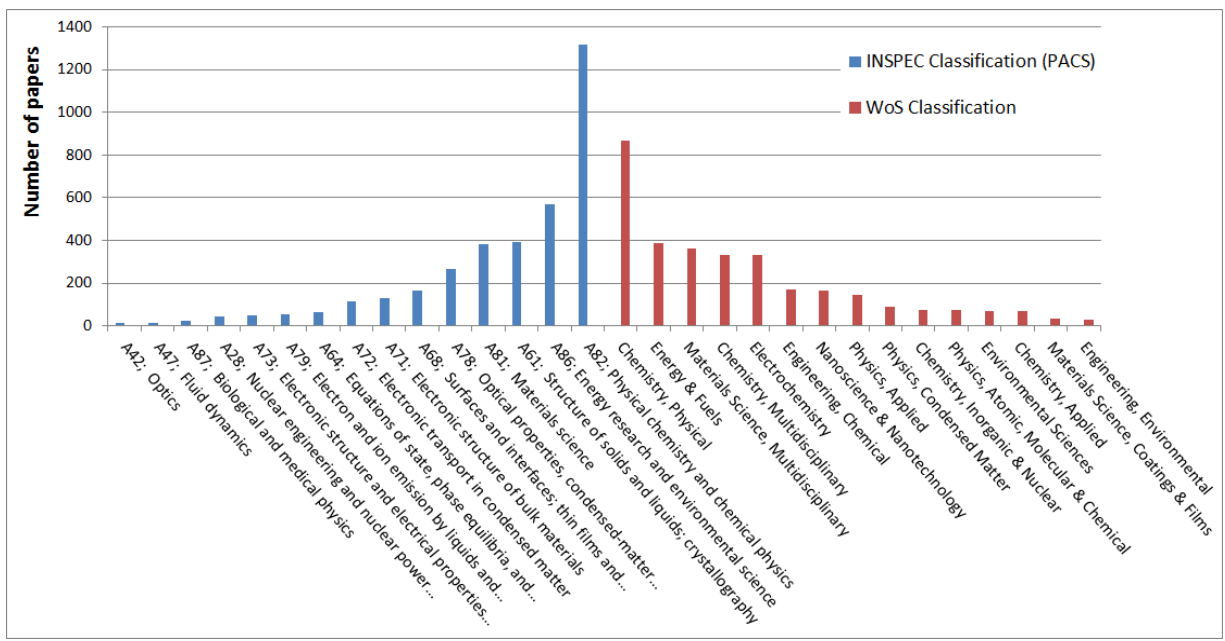

Figure 2: Distribution of the publications found by the WoS topic search (data set 2) across INSPEC and WoS classification systems (in absolute numbers)

The ACCS cluster 3.7.3 contains 1739 papers in total. 695 papers (40.0%) in the ACCS cluster 3.7.3 were also found in our topic search and therefore seem to be relevant for the topic of "overall water splitting". One of the other four labels ("photocatalytic h") is rather redundant to "overall water splitting". The manual inspection of the randomized sample of 123 papers of ACCS cluster 3.7.3 showed (see above) that "bivo4" is used as a precursor or photocatalyst in some studies, catalysts are synthesized via a "solid state reaction method", a "sacrificial agent" is used in the catalytic cycle in some studies. This explains the other labels of ACCS cluster 3.7.3.

**4.3    Comparison of ACCS cluster 3.7.3 with INSPEC and WoS SCs**

Figure 3 shows the distribution of the papers in ACCS cluster 3.7.3 across the INSPEC (n=691 papers) and WoS SCs (n=1739 papers), data set 1. The blue bars present the distribution across the INSPEC SCs and the red bars the distribution across the WoS SCs. The figure shows that the cluster 3.7.3 contains papers which are categorized in very different scientific fields in both, the INSPEC and WoS classification systems (e.g., "Electrochemistry" and "Chemistry, Inorganic & Nuclear" as well as "Structure of solids and liquids; crystallography" and "Optical properties, condensed-matter spectroscopy and other interactions of radiation and particles with condensed matter"). The three highest populated categories are rather multidisciplinary: "Chemistry, Physical", "Materials Science, Multidisciplinary", and "Chemistry, Multidisciplinary" in WoS and "Physical chemistry and chemical physics", "Materials science", and "Structure of solids and liquids; crystallography" in INSPEC. More importantly, the category labels of WoS and INSPEC classifications agree quite well with each other. The WoS and INSPEC categories which are closely related to the topic of overall water splitting ("Energy research and environmental science" in INSPEC and "Energy & Fuels" in WoS) are on the sixth and fifth ranks. In the case of the topic search (cf. Figure 2), these categories appeared on the second rank and were populated with about twice as much publications.

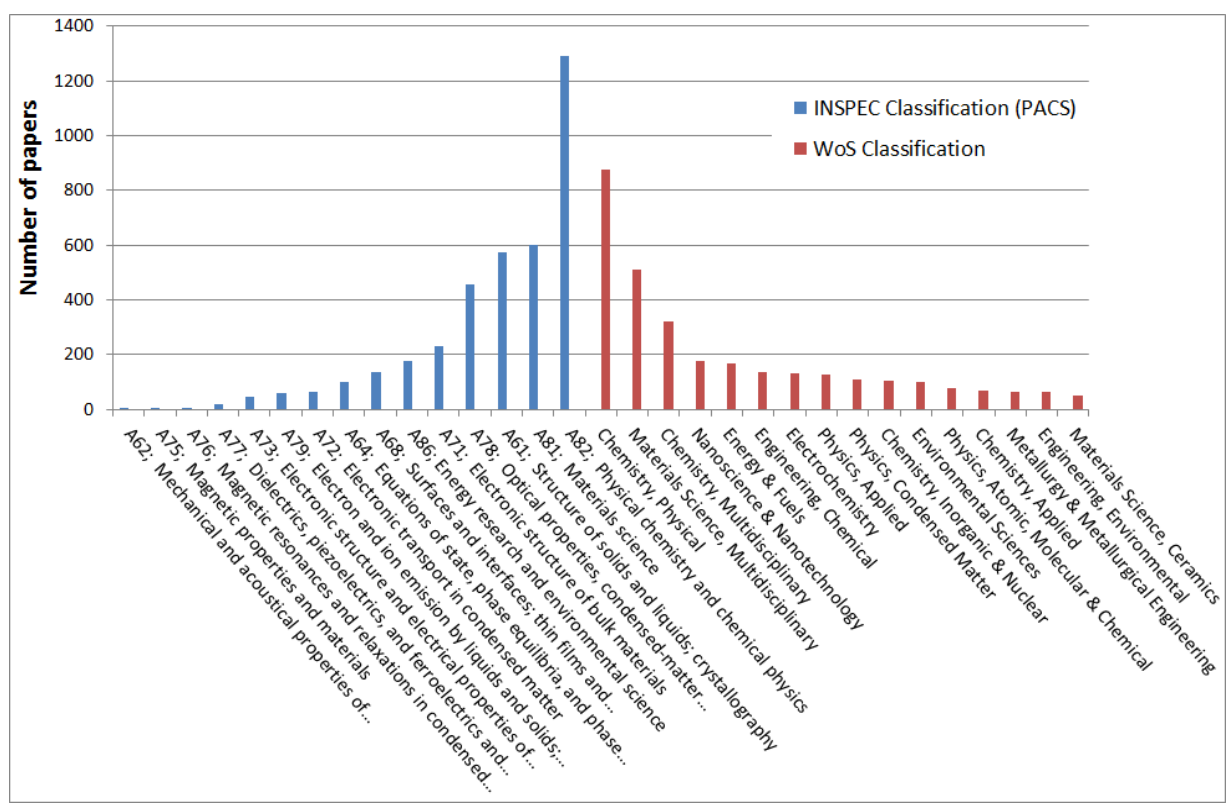

Figure 3: Distribution of the papers in cluster 3.7.3 across the INSPEC and WoS classification systems (in absolute numbers), data set 1

**4.4     Comparison of ACCS cluster 3.7.3 with two related clusters**

We compare three related clusters of similar size ACCS clusters: 3.7.2, 3.7.3, and 3.7.3 (data set 3). The ACCS orders clusters in the same hierarchical level by size (number of papers in the cluster).

Figure 4 shows semantic maps based on a title word analysis of the three clusters. The maps have been produced by using VOSviewer 1.6.5 (www.vosviewer.com). A minimum of 10 occurrences per title word was required in VOSviewer. Although the labels of clusters 3.7.2, 3.7.3, and 3.7.4 are rather different, the title word maps of the papers in these three clusters seem rather similar. For example, Figure 4 shows many terms related to photocatalytic activity and $TiO_2$ in all three clusters as pronounced terms.

However, clusters 3.7.2 and 3.7.3 seem to have more similar characteristics (showing the terms synthesis and preparation) in the semantic maps than cluster 3.7.4 (showing the term degradation more prominently). Judging from the semantic maps, clusters 3.7.2 and 3.7.3 should belong thus to the same cluster (scientific field) and cluster 3.7.4 contains thematically related papers.

cluster 3.7.2

cluster 3.7.3

cluster 3.7.4

Figure 4: Semantic maps of title words of the papers in ACCS clusters 3.7.2, 3.7.3, and 3.7.4. The three maps can be viewed individually via the following URLs: https://tinyurl.com/y8hjpyzv (for ACCS cluster 3.7.2), https://tinyurl.com/y74zl6vw (for ACCS cluster 3.7.3), and https://tinyurl.com/y9lzuuzp (for ACCS cluster 3.7.4).

This pattern is further checked using WoS SCs in Figure 5 where the distribution of the ACCS clusters 3.7.2, 3.7.3, and 3.7.4 across SCs is shown. The figure reveals that the papers in clusters 3.7.2 and 3.7.3 distribute more similarly across the WoS SCs than the papers in cluster 3.7.4. For example, about 24% of the papers in cluster 3.7.2 and 27% in cluster 3.7.3 are assigned to "Chemistry, Physical"; the percentage for cluster 3.7.4 is only about 17%. Also, rather large percentages of publications in ACCS cluster 3.7.4 are assigned to the WoS SCs "Engineering, Chemical", "Environmental Sciences", and "Engineering, Environmental" whereas only rather low percentages of publications from ACCS clusters 3.7.2 and 3.7.3 are assigned to these WoS SCs. Thus, the distribution across WoS SCs is consistent with the pattern which we observed in the semantic maps in Figure 4: The ACCS clusters 3.7.2 and 3.7.3 are more similar to each other than to 3.7.4.

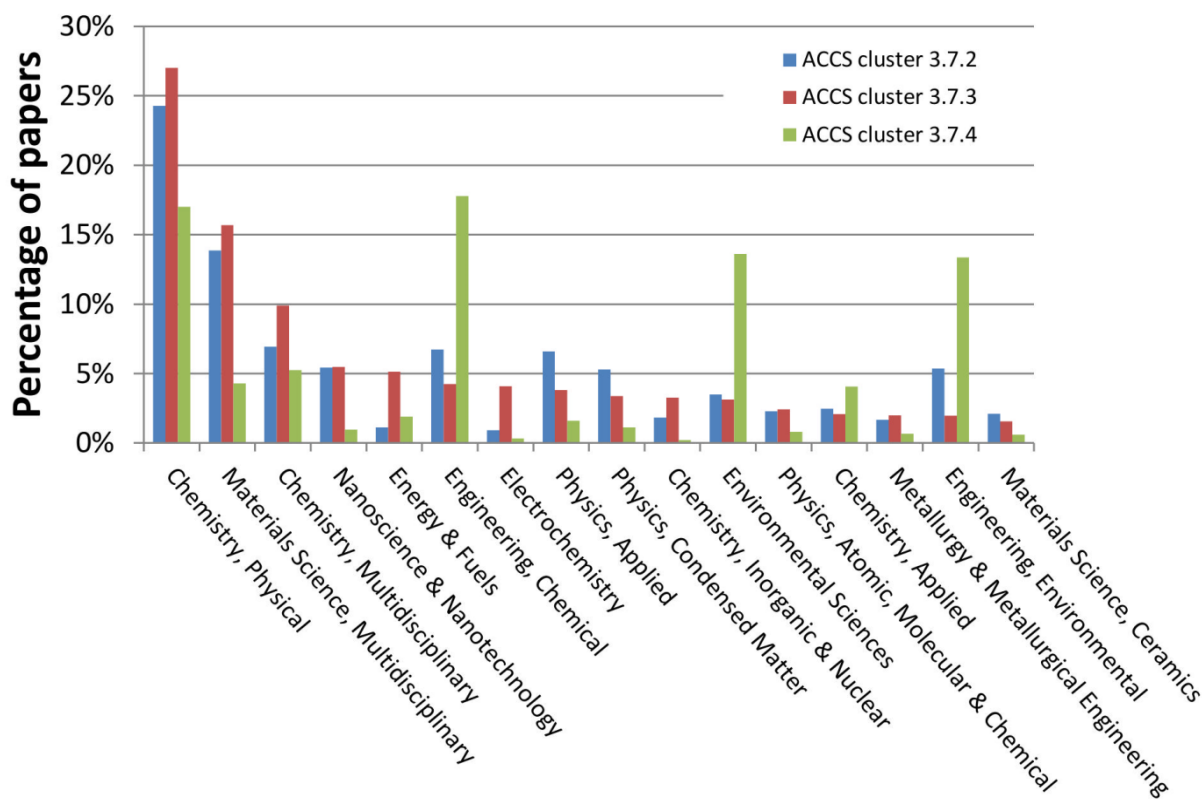

Figure 5: Distribution of the papers in clusters 3.7.2, 3.7.3, and 3.7.4 across WoS SCs (in percentages)

In a final check, we performed a manual evaluation of three randomly selected papers from each of the three clusters. The papers are listed in Table 2.

Table 2: Three randomly selected papers from each of the ACCS clusters 3.7.2, 3.7.3, and 3.7.4.

| Cluster | WoS UT | Title |
|---|---|---|
| 3.7.2 | WOS:000262328000054 | (Bi, C and N) codoped TiO2 nanoparticles |
| 3.7.2 | WOS:000230022700001 | Preparation and characterization of nanosized anatase TiO2 cuboids for photocatalysis |
| 3.7.2 | WOS:000188492100017 | Preparation of a visible light-responsive photocatalyst from a complex of Ti4+ with a nitrogen-containing ligand |
| 3.7.3 | WOS:000279098700015 | One-step hydrothermal coating approach to photocatalytically active oxide composites |
| 3.7.3 | WOS:000245882900006 | A novel scheelite-like structure of BaBi2Mo4O16: Photocatalysis and investigation of the solid solution, BaBi2Mo4-xWxO16 (0.25 <= x <= 1) |

| 3.7.3 | WOS:000230330600021 | Photocatalytic and photophysical properties of visible-light-driven photocatalyst ZnBi12O20 |
|---|---|---|
| 3.7.4 | WOS:000278056300014 | Potential dissolution and photo-dissolution of ZnO thin films |
| 3.7.4 | WOS:000226570200006 | Degradation of methamidophos on soultanina grapes on the vines and during refrigerated storage |
| 3.7.4 | WOS:000226685600022 | Turbidity-based monitoring of particle concentrations and flocculant requirement in wastewater pre-treatment |

All six papers from cluster 3.7.2 and 3.7.3 deal with photocatalytic activity of compounds, their synthesis and study of their physico-chemical properties. Assuming a semantic or topical logic, one would expect to find these six papers either in a single cluster when a broad classification system is used or in many different clusters (one for synthesis and preparation, one for structural analysis, and many other clusters for other analytic methods and properties) with a fine classification system. Two of the three papers from cluster 3.7.4 deal more with photodegradation of compounds and the impacts on fruits and technical products. The third paper from cluster 3.7.4 studies wastewater pre-treatment. Thus, it is surprising to find these three papers in the same cluster. Depending on the granularity of the classification system, one would expect three different clusters (fine granularity) or the two papers about photo degradation in the same cluster and the paper about wastewater pre-treatment in a separate cluster (coarse granularity).

## 4.5 Differences in the average number of citations

The greater thematic similarity of clusters 3.7.2 and 3.7.3 compared to 3.7.4 is also reflected in the average number of citations of the papers, which are presented in Table 3. Table 3 summarizes the number of papers and the average number of citations for the three investigated ACCS clusters. The difference in citation counts between clusters 3.7.2 and 3.7.3 is 6.82 (=52.91-46.09) compared to (1) 13.08 for clusters 3.7.3 and 3.74 (=46.09-33.01) and (2) 19.9 for clusters 3.7.2 and 3.74 (=52.91-33.01) is rather significant. Taken together, the results from Figure 5 and Tables 2 and 3 can be interpreted as follows: although two clusters are more similar to one another than the third cluster to both, the results are contrary to algorithmically constructed fields with high discriminatory power.

Table 3: Number of papers, average number of citations, and expected average number of citations for the three studied ACCS clusters

| ACCS cluster | Number of papers | Average number of citations | Expected average number of citations according to WoS SC |
|---|---|---|---|
| 3.7.2 | 2645 | 52.91 | 16.94 |
| 3.7.3 | 1739 | 46.09 | 17.56 |
| 3.7.4 | 1677 | 33.01 | 16.30 |

Figure 6 shows the average number of citations for all papers assigned to the WoS SCs from Figure 5. It is based on all papers published between 2001 and 2010. The WoS SCs in Figure 6 are ordered by the number of papers also assigned to ACCS cluster 3.7.3 as in Figure 5. We would like to compare these citation counts with the average number of citations in Table 2. The difference between both figures is that the citation counts in Figure 5 are restricted to the papers belonging to the three clusters and those in Figure 6 refer to all papers in the SCs. Citation impact on a similar level would reveal similarities between WoS SCs and the three clusters.

As the results in Figure 6 reveal, however, the average number of citations of the papers in the WoS SCs vary between 5 and 25 citations per paper whereas the average number of citations in Table 2 for the clusters 3.7.2, 3.7.3, and 3.7.4 are 33.01, 46.09, and 52.9 citations per paper, respectively. Obviously, the clusters contain papers with significantly higher average citation counts than the papers in the corresponding WoS SCs.

A comparison of the average number of citations in the ACCS clusters with the WoS SCs may be misleading because each WoS SC is given equal weight although the distribution across the SCs is heterogeneous. Therefore, we perform another comparison of the citation counts in Table 3: The average number of citations can be compared to the expected number of citations. We calculated the expected number of citations based on the average citation count of the WoS SC and the number of papers assigned to this SC. The expected citation count differs insignificantly between the three ACCS clusters 3.7.2, 3.7.3, and 3.7.4. However, large differences are observed between the actual and expected average number of citations: The actual average number of citations is twice as large as the expected one in the case of ACCS cluster 3.7.4. This ratio grows in the case of ACCS cluster 3.7.3 to 2.6 and in the case of ACCS cluster 3.7.2 to 3.1. This substantiates our previous observation that the ACCS clusters contain papers with a higher average citation count than the corresponding WoS SCs.

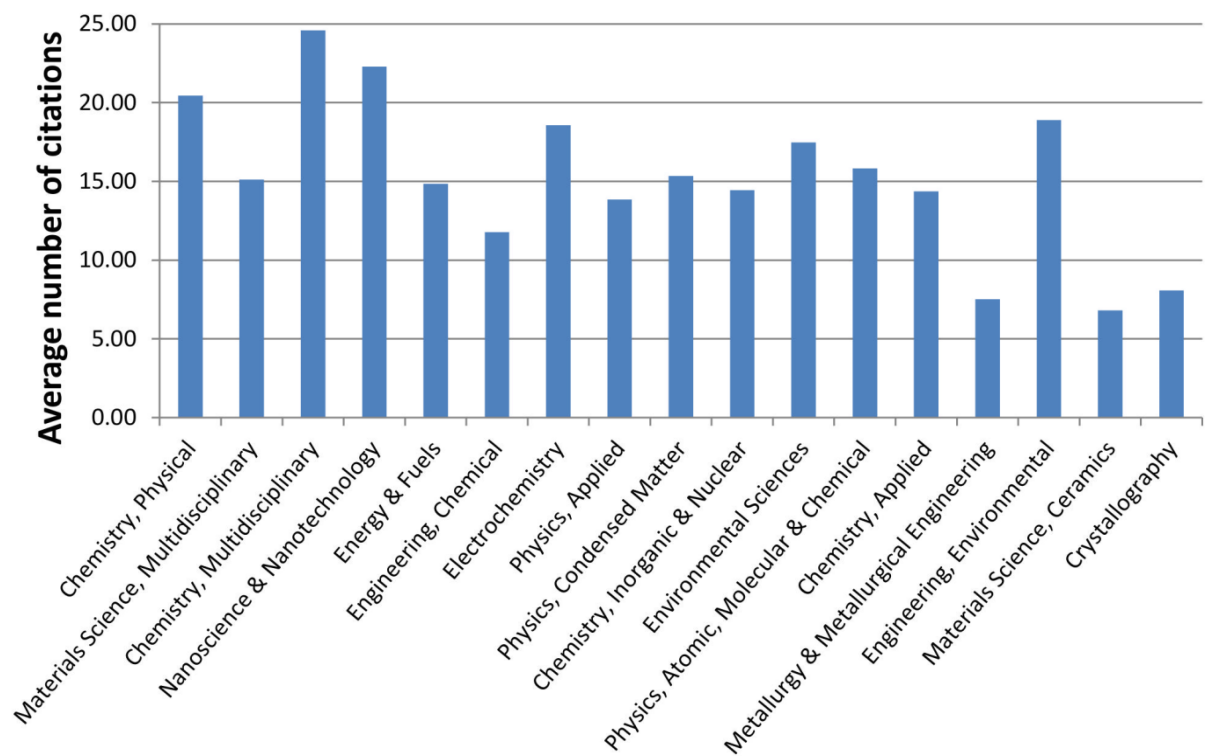

Figure 6: Average number of citations for the WoS SCs from Figure 5 to which the papers in clusters 3.7.2, 3.7.3, and 3.7.4 are assigned

# 5 Discussion

Using papers on the topic "overall water splitting", we have compared the ACCS with two other classification systems (PACS and WoS). Our study follows the recommendation of Waltman and van Eck (2012) for doing such studies: "Another approach may be to compare the results of our methodology with existing publication-level classification systems such as the Physics and Astronomy Classification Scheme (PACS)" (p. 2390). We started with the independent rating by two chemists of a random sample (n=123) of the 1739 publications in the cluster 3.7.3 which has the label "overall water splitting" (besides other labels). 57 (46.3%) of the 123 publications were assigned to the topic of "overall water splitting" by both chemists. 48 (39.0%) of the publications were identified to deal with different topics. In the cases of the remaining 16 publications, the assessments of the chemists disagreed. In a second analysis, we continued with all papers found using a WoS topic search for the field of "overall water splitting". The results show that about 41% of the papers are assigned to the ACCS cluster 3.7.3 which has the label "overall water splitting" (besides other labels). Many papers are spread across 247 other ACCS clusters. In a third analysis, we used all papers in the ACCS cluster which has "overall water splitting" as one label. The results reveal that these papers are assigned to many different WoS and INSPEC categories. The further comparison of cluster 3.7.3 with its related clusters of similar size clusters on the basis of semantic maps and citation counts questions the discriminatory power of the ACCS.

This is the first evaluation study on ACCS, which includes field-specific experts in the analyses. Three of the four co-authors have a chemical and physical background and they are located at a research institute with several researchers in related areas. Another physicist from this research institute has performed an internal review before submission of the manuscript. We think that it is very important and necessary to consider field expertise in the evaluation of (algorithmically constructed) field-classification systems. These systems propose clusters of publications, which are proposed to be related in the reported research (measured by citation relations). Although bibliometricians can produce these systems using the available methods in their area of expertise, they cannot assess based on the content of the clustered publications whether the cluster assignments are reliable and valid. In this study, we selected a micro-level field for which expertise is available by three co-authors and in their wider institutional environment. We would like to encourage experts in other fields to study in more detail cluster solutions of the ACCS, which they are able to assess.

One possible interpretation of our results is that the cluster algorithm used to construct ACCS is not able to distinguish properly between scientific fields. This interpretation is confirmed by similar results which have been published by Rafols and Leydesdorff (2009). They compared content-based and algorithmic classifications of journals (Leydesdorff & Milojević, 2015). However, one should have in mind in the comparison of field classification systems that "the idea of science being subdivided into a number of clearly delineated fields is artificial. In reality, boundaries between fields tend to be rather fuzzy" (Waltman & van Eck, 2013, p. 700). Thus, a completely satisfying solution seems to be impossible. Although the results of this study and Rafols and Leydesdorff (2009) point out that caution should be exercised when normalized indicators based on the algorithmic classification system are used, the results of Perianes-Rodriguez and Ruiz-Castillo (2015) for the Leiden Ranking show that normalized impact values for universities are highly correlated if they are calculated using the WoS SCs or ACCS, respectively.

We assume that the results by Perianes-Rodriguez and Ruiz-Castillo (2015) are mainly due to the aggregation level of universities included in the Leiden Ranking. The results of our study indicate that changes in the field classification system affect the mean citation impact significantly. The normalized impact indicators for a paper (or paper set) are often quite different when they are calculated using different classification schemes. The mean citations, which we calculated for WoS SCs and ACCS clusters are so different that they will result in different normalized impact scores if used as reference sets. Since there is no preference in bibliometrics for one or another classification system, both are equally in use. It is an advantage of the ACCS that it makes the work of bibliometricians easier, because it contains no multiple classifications of papers. The disadvantage is that research evaluation based on ACCS is not transparent because clusters cannot be labelled properly and can contain (depending on the cluster resolution) too many different research fields in a single cluster or research fields are split artificially into different clusters. The problem of assigning proper labels to clusters has already been recognized by Waltman and van Eck (2012) and has been substantiated by our case study.

Since this micro study is based on the papers on only one topic, it is unknown if the results can be generalized. However, one example is enough to point to a general flaw or weakness in an algorithm (Popper, 1961). However, this does not mean that the algorithm will fail in all other possible cases. Further studies should follow with comparisons on other subject-specific databases with broad coverages of related subject areas and preferably intellectual assignments of scientific fields to publications. Future studies should consider as many available proposals as possible in the bibliometric literature for field delineations. An overview can be found in Wouters, et al. (2015).

# Acknowledgements

The bibliometric data used in this paper are from an in-house database developed and maintained by the Max Planck Digital Library (MPDL, Munich) and derived from the Science Citation Index Expanded (SCI-E), Social Sciences Citation Index (SSCI), Arts and Humanities Citation Index (AHCI) provided by Clarivate Analytics (Philadelphia, Pennsylvania, USA). We would like to thank the Centre for Science and Technology Studies (CWTS) for making their ACCS assignments to WoS UTs available. Helpful comments from Ludo Waltman and colleagues from the Max Planck Institute for Solid State Research on an earlier draft of the manuscript are greatly appreciated. The authors thank Michael Eppard for an internal review of the manuscript and Stasa Milojevic as Associate Editor who handled the manuscript for very helpful comments during the review procedure.